\documentclass[12pt,a4paper]{article}

\usepackage{graphicx} 
\usepackage[colorlinks]{hyperref}

\newcommand{\erm}{{\mathrm e}}
\newcommand{\grm}{{\mathrm g}}
\newcommand{\prm}{{\mathrm p}}
\newcommand{\qrm}{{\mathrm q}}
\newcommand{\qbar}{\overline{\mathrm q}}
\newcommand{\Hrm}{{\mathrm H}}
\newcommand{\Wrm}{{\mathrm W}}
\newcommand{\Zrm}{{\mathrm Z}}

\newenvironment{Itemize}{\begin{list}{$\bullet$}%
{\setlength{\topsep}{0.2mm}\setlength{\partopsep}{0.2mm}%
\setlength{\itemsep}{0.2mm}\setlength{\parsep}{0.2mm}}}%
{\end{list}}
\newcounter{enumct}

\newlength{\abstwidth}
\begin{document}
\sloppy
 
\pagestyle{empty}
 
\begin{flushright}
LU TP 16-47\\
MCnet-16-36\\
August 2016
\end{flushright}

\vspace{\fill}

\begin{center}
{\LARGE\bf Status and developments of event generators%
\footnote{presented at the Fourth Annual Large Hadron Collider Physics
Conference (LHCP2016),\\13--18 June 2016, Lund, Sweden}}\\[10mm]
{\Large Torbj\"orn Sj\"ostrand}\\[3mm]
{\it Theoretical Particle Physics,
Department of Astronomy and Theoretical Physics,}\\[1mm]
{\it Lund University, S\"olvegatan 14A, SE-223 62 Lund, Sweden}\\[2mm]
\end{center}

\vspace{\fill}

\begin{center}
\begin{minipage}{\abstwidth}
{\bf Abstract}\\[2mm]
Event generators play a crucial role in the exploration
of LHC physics. This presentation summarizes news and plans for the
three general-purpose $\prm\prm$ generators \textsc{Herwig}, 
\textsc{Pythia} and \textsc{Sherpa}, as well as briefer notes on a
few other generators. Common themes, such as the matching and merging
between matrix elements and parton showers, are highlighted. 
Other topics include a historical introduction, from the Lund 
perspective, and comments on the role of MCnet.
\end{minipage}
\end{center}

\vspace{\fill}

\phantom{dummy}

\clearpage

\pagestyle{plain}
\setcounter{page}{1}

\section{Introduction}

Event generators have come to play a key role in particle physics, 
as the bridge that allows direct contact between the theoretical 
idealized world and the experimental reality. But it is a bridge 
that does not come for free; it has to be designed, constructed and 
maintained, and often expanded to meet more demands. In this presentation 
we will describe the status and development trends among the main 
generators for $\prm\prm$ physics, with main emphasis on the last few
years, but with some historical background. A starting platform is 
provided by the 2011 MCnet review article \cite{Buckley:2011ms}, 
but there are also relevant later reviews 
\cite{Skands:2012ts,Seymour:2013ega,Agashe:2014kda, Hoche:2014rga}.
We also refer to Gavin Salam's presentation for the very important
input provided by the explosion of higher-order calculations
\cite{Salam:2016xyz}.

\begin{figure}[t]
\includegraphics[width=\textwidth]{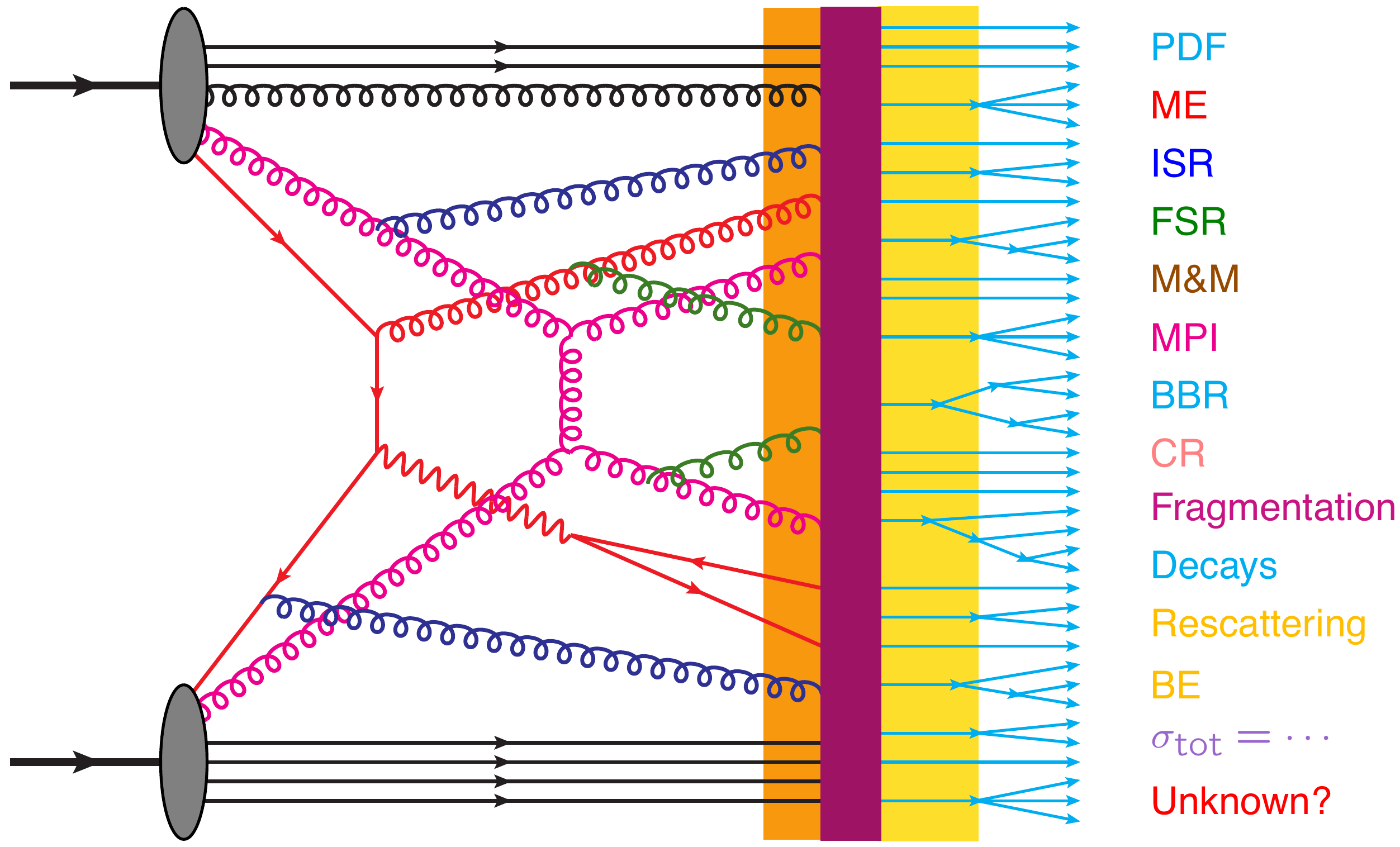}
\caption{The main components in the generation of an event.}
\label{fig1}
\end{figure}

To illustrate the diversity of challenges that a full-fledged generator
has to face, let us begin with a short overview of the main physics 
components, figure~\ref{fig1}, most of which will be revisited. 
\begin{Itemize}
\item Initially two hadrons are coming in on a collision course,
each with a partonic flux given by the Parton Distribution Functions
(PDFs).
\item A collision between two partons, one from each side, gives the
hard process of interest, characterized by the relevant matrix elements
(MEs). These usually are leading order (LO) or next-to-LO (NLO),
increasingly also next-to-NLO (NNLO). 
\item When short-lived ``resonances'' are produced in the hard process, 
such as the top, $\Wrm^{\pm}$, $\Zrm^0$ or $\Hrm^0$, their decay has to be 
viewed as part of this process itself.
\item A collision implies accelerated charges, and thereby bremsstrahlung. 
This can be described in terms of Parton Showers (PSs), usually split
into Initial-State Radiation (ISR) and Final-State Radiation (FSR). 
\item Since PSs describe the same physics as higher-order MEs do, they 
must be combined consistently, with PSs adding multiple softer emissions 
not covered by the fixed-order hard MEs, without gaps or overlaps. This 
is called matching and merging (M\&M), where matching is the procedure 
to obtain a smooth transition for a fixed parton multiplicity and 
merging the combination of several multiplicities, but often the two 
aspects are intertwined.
\item Since the hadrons are made up of a multitude of partons, further 
parton pairs may collide within one single hadron--hadron collision --- 
MultiParton Interactions (MPIs), each associated with its ISR and FSR.
\item Much of the incoming energy and flavours remain in the beam--beam 
remnants (BBRs), which also carry colours that compensate the colours 
taken away by the colliding partons.
\item Until now, the colour structure has been studied on a local level,
and usually bookkept in the $N_C \to \infty$ limit \cite{'tHooft:1973jz},
where colour assignments are unambiguous.
The overall colour topology may be rearranged relative to the naive picture,
colour reconnection (CR), e.g.\ in a way that reduces the ``free energy'' 
in the colour fields.
\item As the partons created in the previous steps recede from each other, 
confinement forces become significant. These colour fields break up 
into a primary generation of hadrons --- fragmentation. 
(Alternatively hadronization, but this can also be used in a more general 
sense, encompassing most nonperturbative aspects.) 
\item Many of those primary hadrons are unstable and decay further at 
various timescales. 
\item Since many hadrons are produced in close proximity the possibility 
exists that they scatter against each other, before or after decays ---
rescattering.
\item For the same proximity reason, there is the potential for significant 
Bose--Einstein (BE) effects among identical particles.
\item So far we have considered events in the context of a hard interaction,
but all aspects of the total cross section must be modeled, not least
the character of diffractive events as opposed to nondiffractive ones.
\item To these points we should add the unknown, e.g. whether some kind 
of Quark--Gluon Plasma (QGP) is produced also in $\prm\prm$ collisions, 
as hinted by recent experimental observations.
\end{Itemize}
The above points are arranged roughly in order of increasing time scales 
and decreasing understanding, as we move from perturbative to 
nonperturbative physics. 

This article is organized as follows. Since 2016 marks the 40th anniversary 
of the Lund QCD phenomenology group, and given the location of this 
conference, section 2 contains a few historical remarks. Section 3 
introduces the three general-purpose generators and the role of MCnet,
while sections 4--6 summarizes news on \textsc{Herwig}, \textsc{Sherpa}
and \textsc{Pythia}, with some common themes in section 7. A brief 
overview of some other generators are presented in section 8, and finally
section 9 contains a summary and outlook. 

\section{Lund phenomenology flashback}

In 1976, when Bo Andersson and G\"osta Gustafson started the Lund group, QCD 
was still young and poorly understood. The idea of a linear confinement 
by some stringlike mechanism existed since the late sixties, and linear 
confinement also had some support in QCD studies. Therefore a straight 
string (without troublesome transverse vibrations) stretched between a 
pair of high-energy partons was chosen as a simple and Lorentz-covariant 
description of a linear potential $V(r) = \kappa \, r$. The string tension 
$\kappa \approx 1$~GeV/fm was known from Regge phenomenology. 
The one-dimensional string is only to be viewed as a simple 
parametrization of a flux tube with a transverse size of hadronic
dimensions, maybe with a structure analogous to a vortex line in a 
type II superconductor.
 
\begin{figure}[t]
\includegraphics[width=0.53\textwidth]{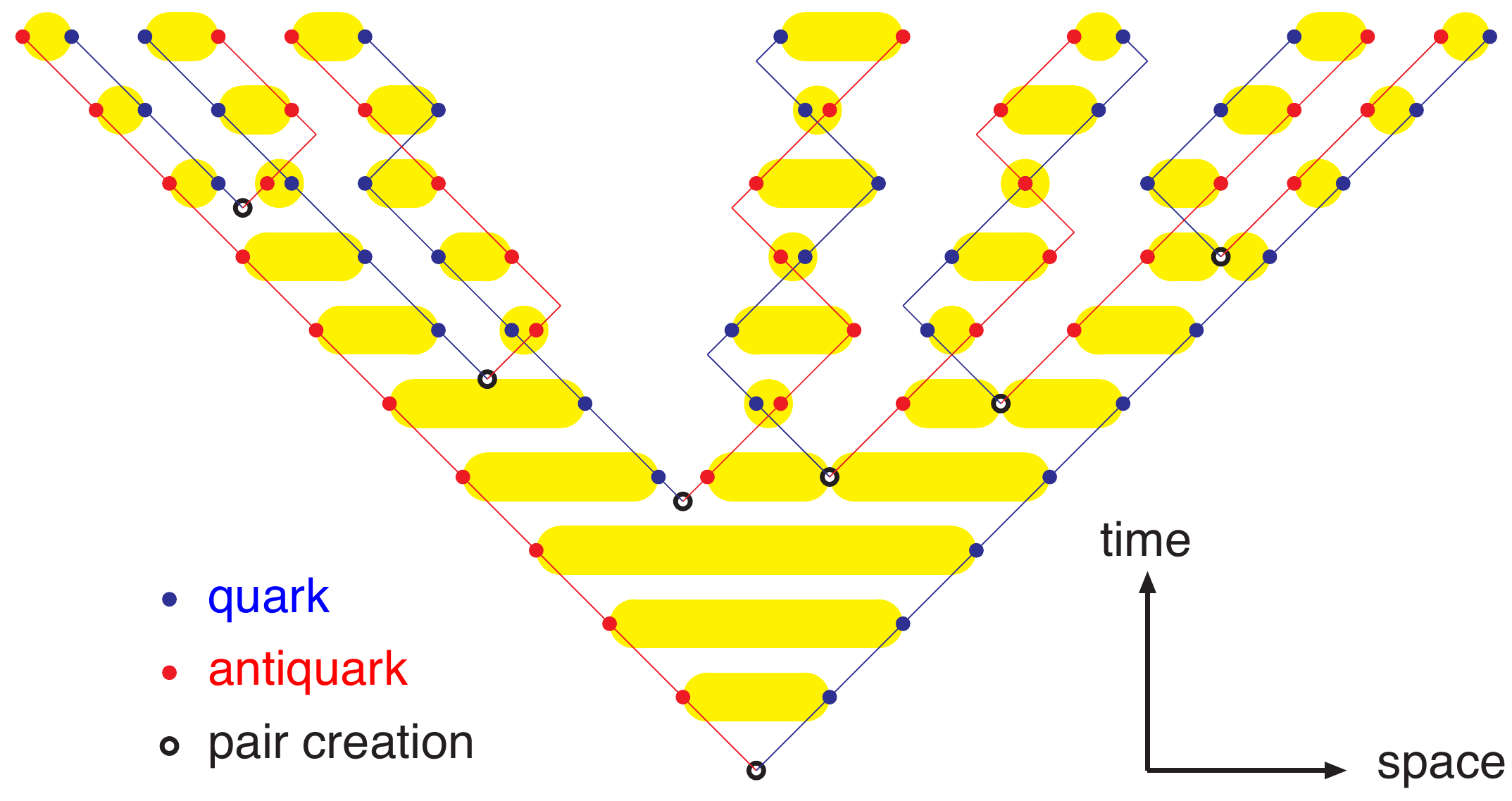}\hfill%
\includegraphics[width=0.4\textwidth]{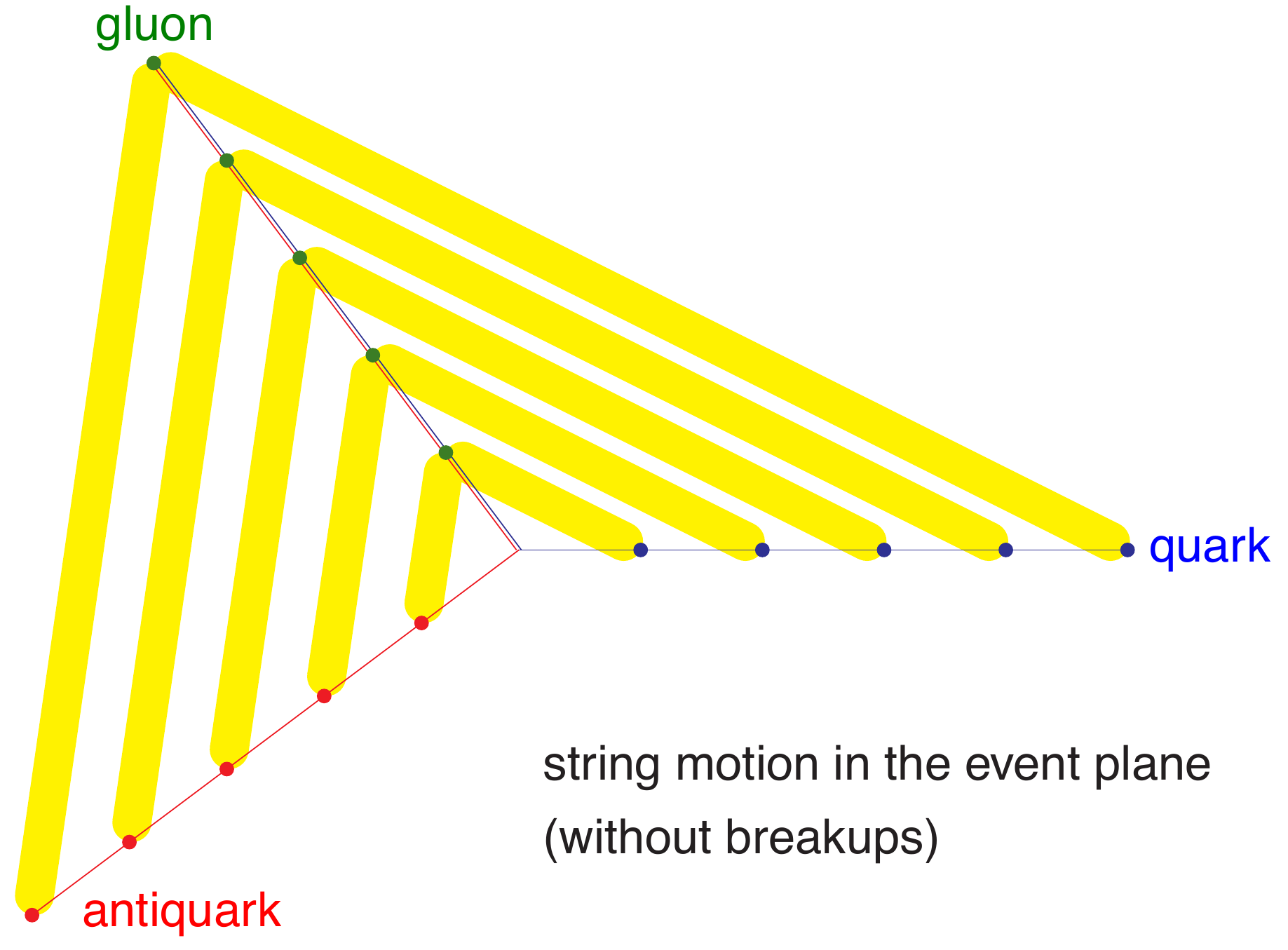}\\
\hspace*{0.25\textwidth}\textit{(a)}\hspace{0.45\textwidth}\textit{(b)}
\caption{\textit{(a)} String breakup in a $\qrm\qbar$ event. 
\textit{(b)} String drawing in a $\qrm\qbar\grm$ event.}
\label{fig2}
\end{figure}

A string stretched between an original $\qrm_0\qbar_0$ pair can break by 
the creation of new $\qrm_i\qbar_i$ pairs, figure~\ref{fig2}\textit{a}.
Ordering them $1 \leq i \leq n - 1$ from the quark end results in the 
production of $n$ hadrons $\qrm_0\qbar_1$, $\qrm_1\qbar_2$, \ldots, 
$\qrm_{n-1}\qbar_0$. 
The fragmentation begins at the center of the event and then spreads 
outwards, roughly along a hyperbola of constant proper time. Since the 
production vertices are spacelike separated, any time ordering is Lorentz
frame dependent, however. It is therefore allowed to consider particle
production e.g. starting at the $\qrm_0$ end, and recursively split the
system into a hadron plus a smaller remainder-system
\cite{Andersson:1978vj}.

The 1978 article by Field and Feynman \cite{Field:1977fa} used computer 
simulation to describe the fragmentation of a single quark. This approach 
allowed more detailed studies than had been possible by previous analytic 
methods, and marked the starting point for a more widespread use of 
computers to provide theory predictions. (A previous study by Artru and 
Mennessier \cite{Artru:1974hr} having passed largely unnoticed.) 
So also in Lund, where the first public version of the \textsc{Jetset} 
program provided a setup that could simulate jet fragmentation both 
according to the Lund prescription and to the Field-Feynman one. It was 
a Fortran program consisting of roughly 200 punched cards, written under 
most primitive conditions, but a beginning. 

In 1980 the fragmentation model was extended from $\qrm\qbar$ to 
$\qrm\qbar\grm$ events. The crucial step was to apply an $N_C \to \infty$ 
picture, wherein one string piece was stretched from the $\qrm$ to the 
$\grm$ and another on from the $\grm$ to the $\qbar$, 
figure~\ref{fig2}\textit{b}. The absence of a corresponding direct 
connection between the $\qrm$ and $\qbar$ led to asymmetries in the 
particle production where, inside the quark jet, e.g., the rate would be 
enhanced (depleted) on the side toward the gluon (antiquark) 
\cite{Andersson:1980vk}. 
These predicted features were rapidly confirmed by the JADE collaboration 
at PETRA \cite{Bartel:1981kh}. The \textsc{Jetset} code had now grown to 
around 1000 lines, but was still sent to DESY as a deck of punched cards. 
Not only did the observation of ``the string effect'' mean a breakthrough 
for the Lund model, but it may also have been the first time when a new 
particle physics idea and the code to help test it were made available 
to the experimental community at the same time. As it happened, the JADE 
study was done somewhat differently than suggested, highlighting the larger 
flexibility of a code relative to an analytic prediction.

This approach has caught on, and the generators of today are very much 
correlated with the development of new physics ideas, be it for parton 
showers, for matching and merging, for multiparton interactions and colour 
reconnection, or for hadronization. Conversely, the user choice of a 
specific generator in a specific setup is an active choice of physics 
ideology, a point that is not always fully appreciated today, where 
technical accuracy (N$^n$LO) is at the center of attention. Since often 
it is not known which is the ``right'' answer, it is necessary to compare 
programs and options, and relate observed differences to the underlying 
physics assumptions. 

Over the years, the Lund group has contributed with a number of further 
new ideas, notably on topics such as 
\begin{Itemize}
\item dipole showers, 
\item backwards evolution for ISR,
\item multiparton interactions,
\item colour reconnection,
\item the modelling of heavy-ion collisions in the absence of QGP formation,
\item matching and merging between matrix elements and parton showers,
\item descriptions of small-$x$ evolution, and
\item handling of QCD effects in many scenarios for physics Beyond the 
Standard Model (BSM).
\end{Itemize}

The generator work to match these physics ideas has been continued, 
most notably with \textsc{Pythia}, which in time subsumed \textsc{Jetset},
but also with \textsc{Fritiof}, \textsc{Ariadne}, \textsc{LDC} and 
\textsc{Dipsy}. Former Lund members have written programs like 
\textsc{Lepto}, \textsc{Vincia} and \textsc{Dire}. And many more programs
make use of \textsc{Pythia}, such as \textsc{RapGap}, \textsc{Hijing}
and even \textsc{Geant}. 

\section{The workhorses and MCnet}

There are three general-purpose event generators for LHC $\prm\prm$ physics:
\textsc{Herwig}, \textsc{Pythia} and \textsc{Sherpa}. They all offer 
convenient frameworks for a wide range of physics studies, covering 
essentially all of the aspects listed in the Introduction, but with 
slightly different historical background and interests.
\begin{Itemize}
\item \textsc{Pythia} goes back to \textsc{Jetset}, begun in 1978,
in the hadronization studies of the time, and further development has 
continued to stress nonperturbative physics.
\item \textsc{Herwig} is the successor to \textsc{Earwig}, begun in 1984,
and originated in the modelling of coherence effects by an 
angular-ordered parton shower, which then was complemented by 
a cluster fragmentation model.
\item \textsc{Sherpa} grew out of a matrix element calculator 
(\textsc{Amegic++}) and a parton shower program (\textsc{Apacic++}), 
begun in 1999, and from the onset stressed the need to match and merge 
these two techniques consistently.
\end{Itemize}
\textsc{Sherpa} is the only of the three to have been written in C++ 
from the onset; the other two have had to undergo a time-consuming
transition from the original Fortran codes.

While competitors at some level, the three projects also benefit from 
each other, and join forces within the EU-funded MCnet network. This 
network also encompasses a number of other projects, notably the 
\textsc{MadGraph} matrix element generator, but also the \textsc{Ariadne}, 
\textsc{Dipsy} and \textsc{HEJ} ``plugin'' programs to the main generators, 
and standard general facilities such as \textsc{Rivet} \cite{Buckley:2010ar}, 
\textsc{Professor} \cite{Buckley:2009bj}, \textsc{LhaPdf} 
\cite{Buckley:2014ana}, \textsc{HepMC} \cite{Dobbs:2001ck}
and HepForge \cite{Buckley:2006nm}. In addition to 
internal activities, like generator development and PhD student training,
it also offers services to the community. One such is the possibility for 
PhD students to come to a node for 3--6 months, to carry out some specific 
projects of mutual interests. Another is the arrangements of summer schools 
on event generators, in 2016 at DESY together with CTEQ, in 2017 in Lund,
3--7 July. Both experimentalists and theorists are most welcome! 

MCnet has received funding 2007--10, 2013--16, and is now approved for 
2017--20. Nodes in the new incarnation are Manchester, Durham, 
Glasgow, G\"ottingen, Heidelberg, Karls\-ruhe, UC London, Louvain 
and Lund, with CERN, SLAC and Monash (Melbourne) as partners.

\begin{table}[t]
\begin{tabular}{|l|l|}
\hline
Program & Homepage \\ \hline
\textsc{Herwig} & \href{https://herwig.hepforge.org/}%
{\texttt{https://herwig.hepforge.org/}}\\
\textsc{Pythia} & \href{http://home.thep.lu.se/Pythia/}%
{\texttt{http://home.thep.lu.se/Pythia/}} \\
\textsc{Sherpa} & \href{https://sherpa.hepforge.org/trac/wiki}%
{\texttt{https://sherpa.hepforge.org/trac/wiki}} \\
\hline
\textsc{Dipsy} & \href{http://home.thep.lu.se/DIPSY/}%
{\texttt{http://home.thep.lu.se/DIPSY/}} \\
\textsc{Dire} & \href{https://direforpythia.hepforge.org/}%
{\texttt{https://direforpythia.hepforge.org/}} \\
\textsc{HEJ} &  \href{http://hej.web.cern.ch/HEJ/index.html}%
{\texttt{http://hej.web.cern.ch/HEJ/index.html}} \\
\textsc{Vincia} & \href{http://vincia.hepforge.org/}%
{\texttt{http://vincia.hepforge.org/}} \\
\hline
\end{tabular}
\caption{Home pages of some of the generators mentioned in this article.}
\label{tab1}
\end{table}

\section{\textsc{Herwig} news}

\textsc{Herwig}~7.0 was released at the end of 2015 \cite{Bellm:2015jjp}.
This is a continuation of the \textsc{Herwig++} development series
\cite{Bahr:2008pv}, begun 16 years earlier, to replace the Fortran 
\textsc{Herwig}~6 program with one in C++. This objective is now achieved; 
the \textsc{Herwig++}~3.0 code fully supersedes the \textsc{Herwig}~6 one, 
and was therefore renamed \textsc{Herwig}~7.0. It replaces all previous 
\textsc{Herwig} and \textsc{Herwig++} versions.

\begin{figure}[t]
\hspace*{0.1\textwidth}\includegraphics[width=0.8\textwidth]{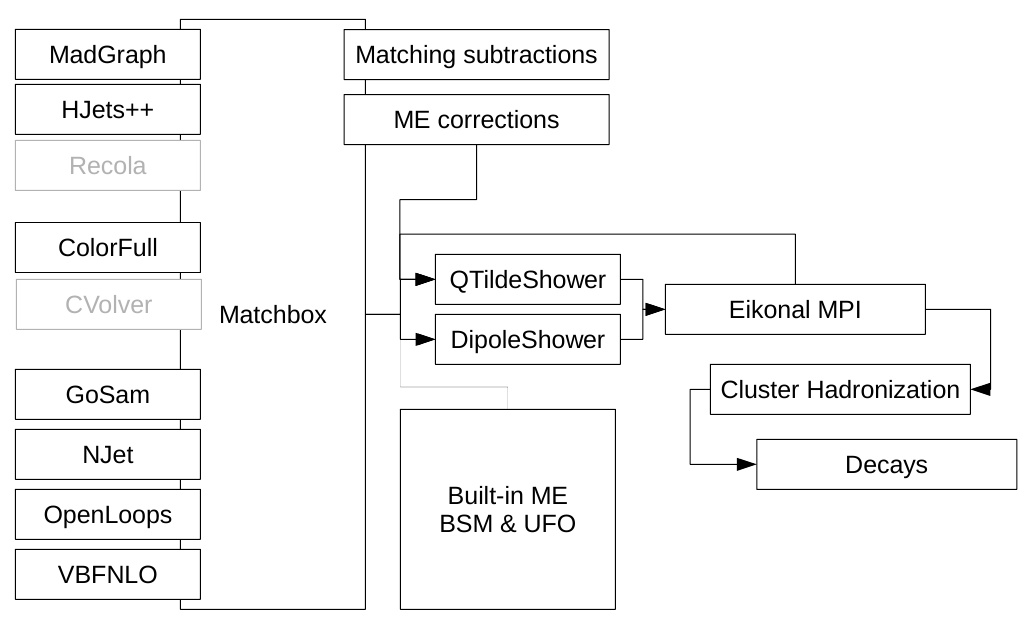}
\caption{The structure of the \textsc{Herwig}~7 program, highlighting 
the role of \textsc{Matchbox} \cite{Plaetzer:2016xyz}.}
\label{fig3}
\end{figure}

The main physics news is that NLO matched to parton showers is default
for most SM processes. This is fully automated, with no external codes
to run separately and no intermediate event files. There is a choice between 
a subtractive (MC@NLO type \cite{Frixione:2002ik}) and a multiplicative 
(PowHeg type \cite{Nason:2004rx}) matching. This is made possible by
the \textsc{Matchbox} module \cite{Platzer:2011bc}, figure~\ref{fig3}, 
which interfaces to 
a large number of different external providers of tree and one-loop
amplitudes. The NLO matching subtractions are then performed internally, 
using the Catani--Seymour dipole approach \cite{Catani:1996vz}. All 
auxiliary programs, the ME ones as well as general utilities such as 
\textsc{LhaPdf} and \textsc{FastJet} \cite{Cacciari:2011ma}, can be 
downloaded, installed and built together with \textsc{Herwig} with the 
help of a bootstrap script. 

\begin{figure}[b!]
\includegraphics[width=0.5\textwidth]{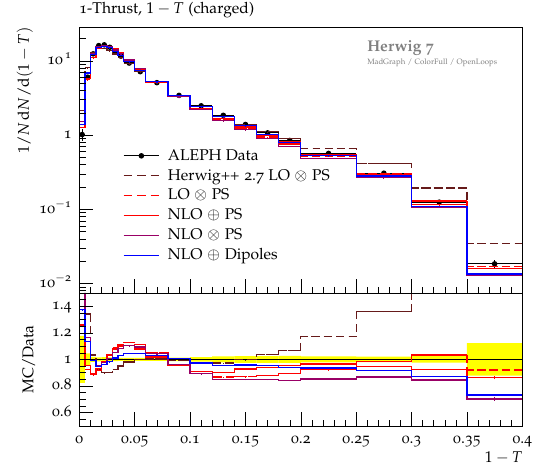}%
\includegraphics[width=0.5\textwidth]{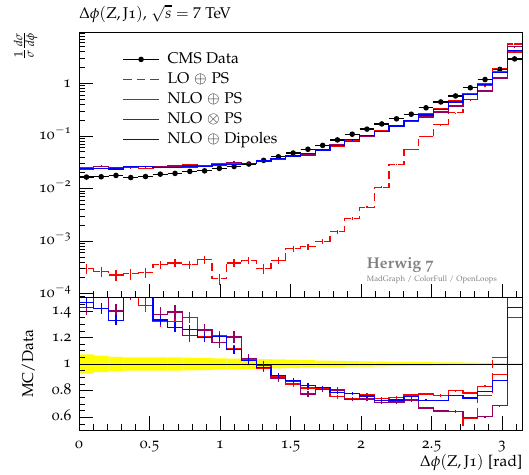}\\
\hspace*{0.25\textwidth}\textit{(a)}\hspace{0.45\textwidth}\textit{(b)}
\caption{Examples of \textsc{Herwig}~7 comparisons with data 
\cite{Bellm:2015jjp} , including the new tune, \textit{(a)} for the 
$1 - T$ distribution in $\erm^+\erm^-$ and \textit{(b)} the azimuthal 
angle between a $\Zrm^0$ and its hardest recoil jet.}
\label{fig4}
\end{figure}

\textsc{Herwig} now comes with two different shower algorithms, the 
traditional angular-ordered one \cite{Gieseke:2003rz} and a newer dipole 
shower \cite{Platzer:2009jq}.  The former has been improved in a number 
of respects, e.g. to include QED showering and spin correlations (at LO), 
and $\grm \to \qrm\qbar$ branchings are no longer angular-ordered to 
reflect the absence of a soft singularity in it. Uncertainties from scale
variations in the shower can now automatically be evaluated, see further 
section~\ref{sect:common}.

Also many other improvements have been made, such as new default tune
that includes both minimum-bias and underlying-event data, 
exemplified in figure~\ref{fig4}, or several options to parallelize 
execution. Last but not least, the documentation of the program has 
been vastly improved, see the web page for the code, manual, further 
instructions and lots of auxiliary material.

In the near future the \textsc{Herwig}~7.1 release will bring 
NLO multijet merging, based on unitarized merging ideas. Other features 
under development include NLO handling of loop-induced processes,
an extended UFO Feynman rules \cite{Degrande:2011ua} support, extended
handling of event reweighting from weight vectors in \textsc{HepMC} files, 
improved top decays in the dipole shower, an interface to the \textsc{HEJ} 
program, and new models for soft interactions and diffraction. 
The already existing code will also be used for numerous physics studies, 
to explore and highlight the new possibilities \cite{Rauch:2016upa}.

Noteworthy is that \textsc{Herwig}~7 has grown to a size of more than 
500,000 lines of code. This, as well as the scope of the code, e.g.\
as manifested by the \textsc{Matchbox} module, has exceeded original
expectations. In the longer run there is therefore a need for a 
significant restructuring, although nowhere near the step between 
\textsc{Herwig}~6 and \textsc{Herwig}~7. There are also many ideas for 
future projects, like amplitude-based parton showers where more 
interference information can be utilized, e.g. subleading colour
contributions \cite{Platzer:2012np}.  

\section{\textsc{Sherpa} news}

\textsc{Sherpa} is undergoing strong but smooth evolution. The latest
published manual, for \textsc{Sherpa} 1.1, is already some years old 
\cite{Gleisberg:2008ta}, but the \textsc{Sherpa} homepage contains an 
extensive manual for the current \textsc{Sherpa}~2.2.1 version,
along with other documentation and source code.

\begin{figure}[t]
\includegraphics[width=\textwidth]{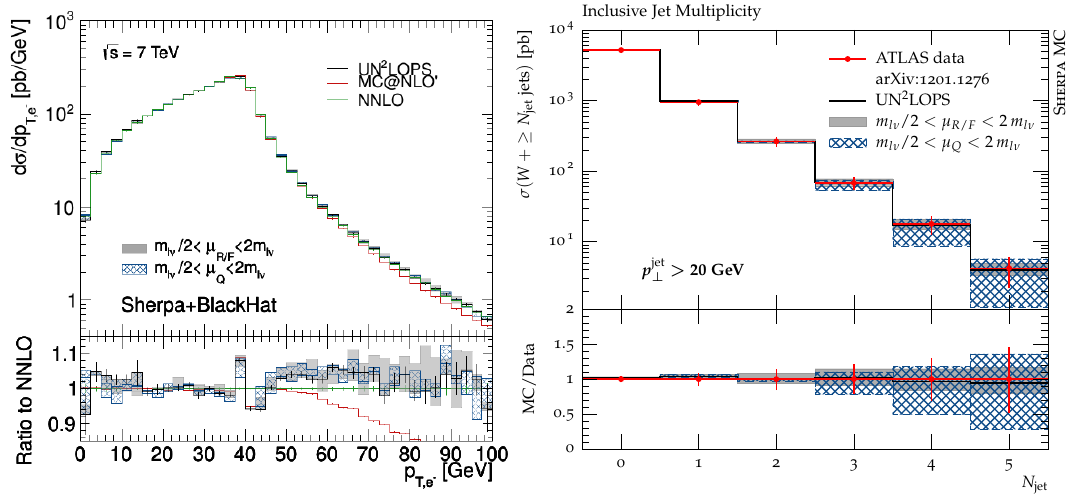}
\hspace*{0.25\textwidth}\textit{(a)}\hspace{0.45\textwidth}\textit{(b)}
\caption{\textsc{Sherpa} results at NNLO+PS for $\Wrm^-$ production.
\textit{(a)} Lepton $p_{\perp}$ spectrum \cite{Hoeche:2014aia}.
\textit{(b)} Inclusive jet multiplicity distribution \cite{Schumann:2016xyz}.}
\label{fig5}
\end{figure}

A significant step is that \textsc{Sherpa} is the first major generator to 
combine processes at NNLO with parton showers, for $\Wrm^{\pm}$,
$\gamma^*/\Zrm^0$ and $\Hrm^0$ production \cite{Hoeche:2014aia,Hoche:2014dla}. 
Examples of results are found in figure~\ref{fig5}, showing excellent 
agreement with data. One-loop matrix elements were obtained from the 
\textsc{BlackHat} library \cite{Berger:2008sj}, and different 
multiplicities were combined using the new UN$^2$LOPS scheme, wherein 
unitarity arguments play a key role.

A new dipole shower, \textsc{Dire}, is available as an option to the 
existing one \cite{Schumann:2007mg}. It will be described further in 
section~\ref{sect:other}.

Another new feature is a fast way to evaluate uncertainties from 
the choice of $\alpha_s(M_{\Zrm})$, PDF, renormalization
and factorization scales, and inherent shower algorithm choices
\cite{Bothmann:2016nao}. The shower reweighting will be discussed
in section~\ref{sect:common}. The matrix element part is carried out
at NLO, where real emission and one-loop corrections are separated 
using Catani--Seymour dipole subtraction \cite{Catani:1996vz}.
The \textsc{MCgrid} package \cite{Bothmann:2015dba} allows for fast and 
flexible calculation on interpolation grids.

Electroweak (EW) NLO corrections have been included 
\cite{Kallweit:2014xda,Kallweit:2015dum} for $\Wrm/\Zrm$ production.
These are the first steps towards an automatization of full NLO 
QCD + EW corrections, combined with showers. Another area of study is 
corrections for loop-induced processes.

\begin{figure}[b!]
\includegraphics[width=0.5\textwidth]{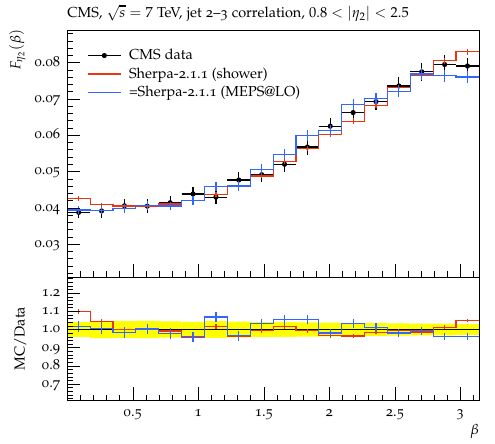}%
\includegraphics[width=0.5\textwidth]{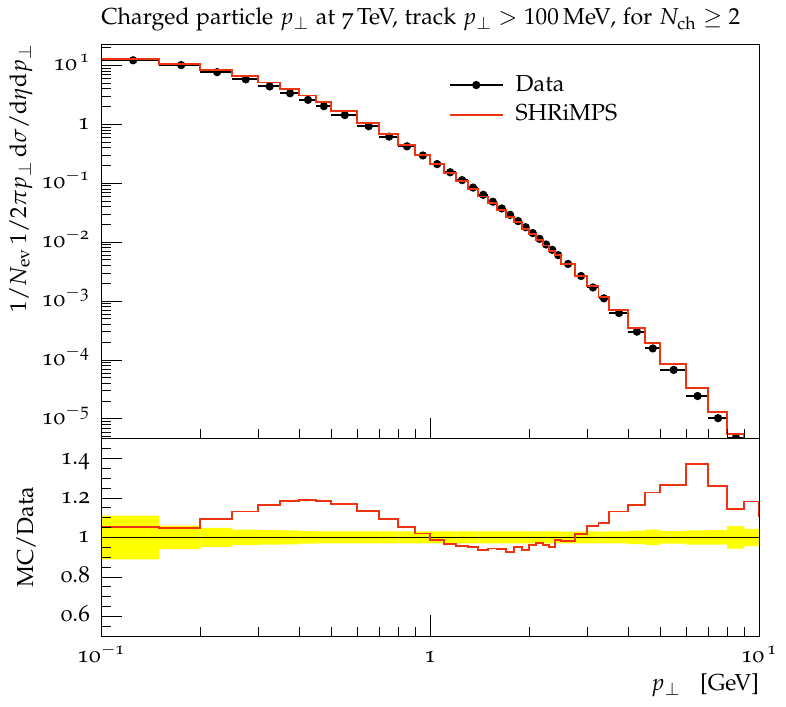}\\
\hspace*{0.25\textwidth}\textit{(a)}\hspace{0.45\textwidth}\textit{(b)}
\caption{\textit{(a)} Coherence angle distribution \cite{Schumann:2016xyz}, 
see \cite{Chatrchyan:2013fha} for definition .
\textit{(b)} Inclusive charged particle $p_{\perp}$ spectrum in the 
\textsc{SHRiMPS} framework \cite{Schulz:2016vml}.}
\label{fig6}
\end{figure}

Several physics studies are steadily performed. As an example,
coherence in QCD events, e.g the angular distribution of a third soft
jet around the second harder jet, is not so well described in 
\textsc{Herwig} and \textsc{Pythia} \cite{Chatrchyan:2013fha}, but
comes out very well for \textsc{Sherpa} already from the pure shower,
figure~\ref{fig6}\textit{a}.
 
The current \textsc{Pythia}-inspired MPI framework will eventually be
replaced by the \textsc{SHRiMPS} implementation of the 
Khoze--Martin--Ryskin model \cite{Ryskin:2009tj}.
A tuning effort is underway \cite{Schulz:2016vml}, and the current
status is exemplified in figure~\ref{fig6}\textit{b}.

The key aim of ongoing work is to achieve full NNLO QCD + NLO EW accuracy,
combined with showers. These showers will go beyond the current 
leading-log framework, to include one-loop corrections,
which also implies new $1 \to 3$ splitting kernels (or $2 \to 4$ in
dipole language) and subleading colour corrections. There are also
studies to automatize $N$-jettiness slicing \cite{Stewart:2010tn}
to allow more effective generation.

\section{\textsc{Pythia} news}

\textsc{Pythia}~8.2 was released two years ago \cite{Sjostrand:2014zea},
but represented a rather smooth upgrade from 8.1, which however was a major
rewrite from the Fortran-based 6.4 \cite{Sjostrand:2006za}. New subversions 
are released $\sim 3$ times per year, the most recent being 8.219. The 
\textsc{Pythia} homepage contains downloads and full documentation.

The code is distributed with a number of different M\&M methods that 
can operate on external ME input, e.g. from \textsc{MadGraph5\_aMC@NLO} 
\cite{Alwall:2014hca} or the \textsc{PowHeg Box} \cite{Alioli:2010xd},
and match those to the internal parton showers. The newest addition to 
this list is the FxFx method \cite{Frederix:2012ps}.

\begin{figure}[b!]
\includegraphics[width=0.5\textwidth]{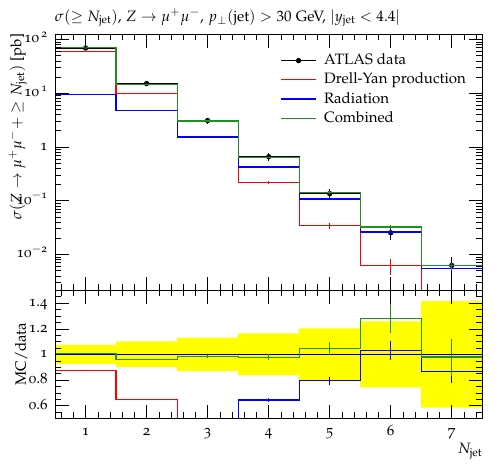}%
\includegraphics[width=0.5\textwidth]{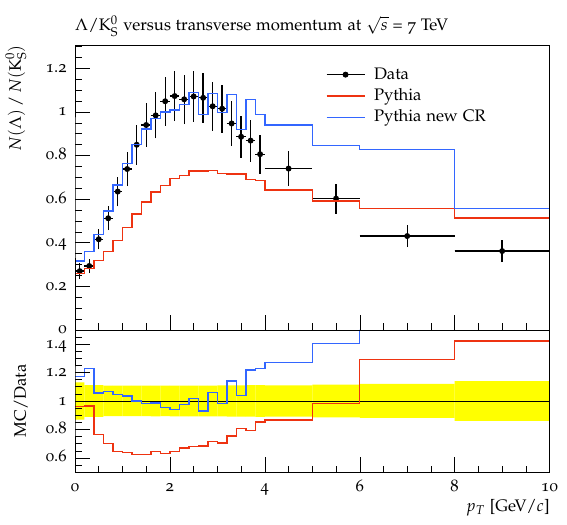}\\
\hspace*{0.25\textwidth}\textit{(a)}\hspace{0.45\textwidth}\textit{(b)}
\caption{\textit{(a)} The inclusive $\Zrm^0 + n$-jet rate
\cite{Christiansen:2014kba}. 
\textit{(b)} The $\Lambda / \mathrm{K}^ 0_S$ production ratio as function 
of $p_{\perp}$ \cite{Christiansen:2015yqa}.}
\label{fig7}
\end{figure}

The shower algorithm has been extended to also include weak branchings,
$\qrm \to \qrm \Zrm^0$ and $\qrm \to \qrm' \Wrm^{\pm}$ 
\cite{Christiansen:2014kba}. The branching kernels are corrected to 
representative $s$- and $t$-channel MEs to obtain a realistic behaviour 
across the full phase space. The well-known shortfall in the 
$\Zrm + n$-jet rate, when only QCD emission off a $\qrm\qbar \to \Zrm^0$ 
base configuration is considered, is well compensated by $\Zrm^0$ 
radiation off a QCD process, figure~\ref{fig7}\textit{a}. The new 
branchings also offer improvements for M\&M strategies 
\cite{Christiansen:2015jpa}.

The shower has also been extended to support scale variation evaluation
and reweighting of rare shower branchings. The interface for external
shower plugins (used e.g.\ by \textsc{Vincia} and \textsc{Dire}) has been
extended, such that the internally implemented M\&M schemes can be used
also by these.

Many new colour reconnection models have been introduced, in addition 
to the default one. These have been used to explore features of normal 
LHC QCD events, e.g. with respect to baryon production properties
\cite{Christiansen:2015yqa}, figure~\ref{fig7}\textit{b}, top mass 
uncertainties \cite{Argyropoulos:2014zoa} and uncertainties in 
measurements of $\Wrm^{\pm}$ and $\Hrm^0$ properties at a future 
$\erm^+\erm^-$ collider \cite{Christiansen:2015yca}.

While not part of \textsc{Pythia} development efforts, it is 
worth noting that a recent article \cite{Butenschoen:2016lpz} makes 
considerable progress in the long-standing problem of relating the 
\textsc{Pythia} top mass to more traditional pole or 
$\overline{\mathrm{MS}}$ definitions.

Among other physics changes, a new model for hard diffraction is 
available \cite{Rasmussen:2015qgr} and double onium production has been
implemented. Several new tunes from the ATLAS and CMS collaborations 
are easily available, but the Monash one \cite{Skands:2014pea} is
default. On the technical side, LHEF~v3 \cite{Andersen:2014efa} is now 
fully supported, it is possible to run the \textsc{MadGraph5\_aMC@NLO} 
and \textsc{PowHeg Box} programs from within \textsc{Pythia}, and
a new interface allows a Python main program to access the full 
\textsc{Pythia} functionality. 

Ongoing work and plans for the future include implementations of 
$\gamma\gamma$, $\gamma\prm$ and $\erm\prm$ collisions, new expressions
for total, elastic and diffractive cross sections, and alternative 
hadronization sche\-mes.

\section{Common themes}
\label{sect:common}

Although the three generators are developed independently, and are 
fully separate codes, there are common trends in their evolution.
This comes both from mutual influences and from evolution in the particle
physics field as a whole. The most obvious example is that matching and 
merging strategies have dominated the drive towards higher accuracy for
the last 15 years or so. An important precondition is the greatly 
increased calculational capabilities, e.g. for automated NLO calculations,
available through a wide selection of programs.
But, starting from this platform, many persons have contributed with 
schemes to combine matrix elements and parton showers as effectively as 
possible. The evolution is still very much under way, and there is not 
one commonly agreed method that does it all. The ongoing debate is very 
intense and healthy, and it would be impossible to cover it in any
detail here. Still, with the risk of doing injustice, here follows a   
very brief layman's summary.

\begin{figure}[t]
\includegraphics[width=\textwidth]{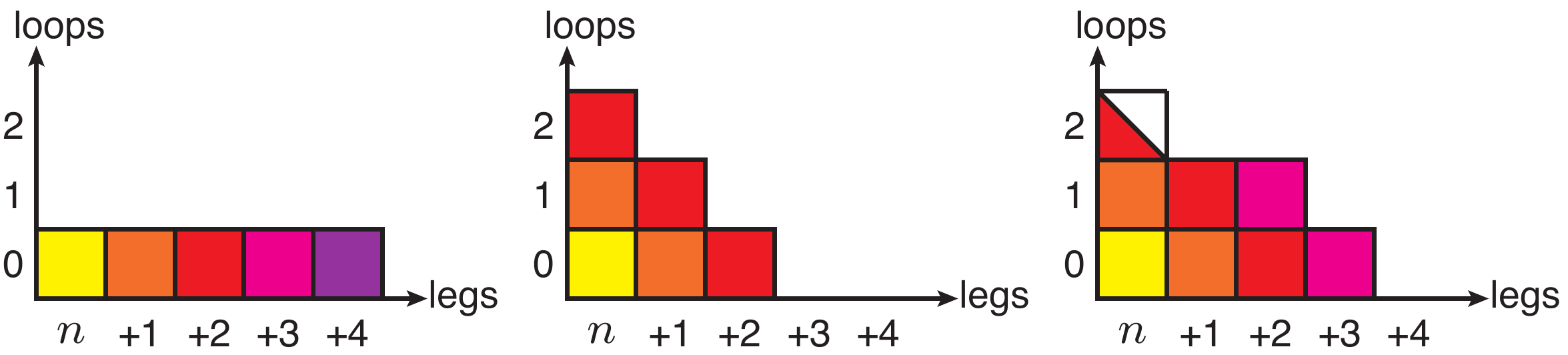}
\hspace*{0.13\textwidth}\textit{(a)}\hspace{0.31\textwidth}\textit{(b)}%
\hspace{0.31\textwidth}\textit{(c)}
\caption{Three main classes of match and merge strategies.
\textit{(a)} Multileg.
\textit{(b)} Order-by-order.
\textit{(c)} Multileg NLO.}
\label{fig8}
\end{figure}

Historically, two different tasks were addressed separately. 
One is multileg merging, figure~\ref{fig8}\textit{a}. The task there 
is to combine LO expressions for different parton multiplicities:
$n$, $n+1$, $n+2$, etc., where $n=0$ for $\Wrm/\Zrm$ production and
$n=2$ for QCD jets. If combined brute-force there will be doublecounting,
since a higher jet multiplicity contributes to the inclusive sample of 
lower multiplicities. With showers added, such doublecounting is further
increased. The solution is to include Sudakov factors, as an 
approximation to higher-order virtual corrections. Or, alternatively put,
as a manifestation of a balance between real and virtual corrections, 
where the addition of an extra jet to an event moves it to another 
jet multiplicity class while keeping the total cross section 
(approximately) unchanged. Approaches of this kind 
include CKKW \cite{Catani:2001cc}, CKKW-L \cite{Lonnblad:2001iq}, 
MLM \cite{Mangano:2006rw} and UMEPS \cite{Lonnblad:2012ng}.     

The other task is to include full higher-order information, to begin
with at NLO, figure~\ref{fig8}\textit{b}. Here the two traditional 
approaches are MC@NLO \cite{Frixione:2002ik} and POWHEG 
\cite{Bengtsson:1986hr,Nason:2004rx}. The key difference is the 
handling of the $n+1$ term, which is only accurate to LO. In MC@NLO it 
is retained as is, while in POWHEG it is rescaled by the same factor 
$K = \sigma_{\mathrm{NLO}}/\sigma_{\mathrm{LO}}$ as the total
cross section. It is possible to interpolate smoothly between these 
options \cite{Alioli:2008tz}. 

The next step along the NLO path is to address NNLO, 
figure~\ref{fig8}\textit{b}, and there are also some schemes
directly aimed at that. The main thrust of recent studies 
has been in a seemingly different direction, however, namely the 
combination of several NLO calculations, the multileg NLO of 
figure~\ref{fig8}\textit{c}. That is, assuming that you have separate 
NLO calculations e.g.\ for $n$ and $n + 1$ partons, how do you combine 
them to achieve NLO accuracy for both $n$ and $n + 1$ topologies and 
LO for $n+2$, also when you match them to showers? As it turns out,
multileg NLO can often comparatively easily be extended to NNLO, by 
adjusting the $n$-parton cross section to preserve a known 
\mbox{(semi-)}inclusive NNLO cross section. Among this broader class of 
algorithms one may note MENLOPS \cite{Hamilton:2010wh},
MINLO \cite{Hamilton:2012np}, 
MINLO$'$ \cite{Hamilton:2012rf,Frederix:2015fyz},
MEPS@NLO \cite{Hoeche:2012yf}, UNLOPS \cite{Lonnblad:2012ix},
FxFx \cite{Frederix:2012ps}, NNLOPS \cite{Hamilton:2013fea}, 
UN$^2$LOPS \cite{Hoeche:2014aia,Hoche:2014dla}, 
GENEVA \cite{Alioli:2013hqa}, and the Pl\"atzer method 
\cite{Platzer:2012bs} further developed for \textsc{Herwig}~7
\cite{Bellm:2016xyz}.

As has already been mentioned, factorization and renormalization scale
variations inside the parton showers have been introduced recently 
both for \textsc{Herwig} \cite{Bellm:2016voq}, \textsc{Sherpa} 
\cite{Bothmann:2016nao} and \textsc{Pythia} \cite{Mrenna:2016sih}. 
The idea is the same: while generating events with the default scale 
choices, construct alternative event weights, e.g. for a grid of scales.
Each weight encodes how likely that shower history would have been,
relative to the default one. By suitable reweighting it is thereby 
feasible to study the scale dependence of any physical observable
without having to regenerate and reanalyze events from scratch.  
It is fairly straightforward to evaluate the part of the total weight
that comes from each individual parton branching. The real trick is that
the effect of the no-emission Sudakov factors can be obtained from 
weight factors associated with those trial emissions that fail, when the 
standard veto algorithm is used to evolve showers.

As a final note, it may be interesting to compare how programs are 
intended to operate in a world with many external sources of ME input.
Here \textsc{Sherpa} is the most restrictive code, that does not 
accept parton-level LHA/LHEF \cite{Alwall:2006yp,Andersen:2014efa} input 
from external ME-level generators. All LO MEs are instead calculated 
internally, while NLO loop corrections are obtained via the BLHA 
interface \cite{Binoth:2010xt} or through dedicated interfaces, 
e.g.\ to \textsc{OpenLoops} \cite{Cascioli:2011va}. It is also possible 
to provide new ME rules \cite{Hoche:2014kca} via either UFO 
\cite{Degrande:2011ua} or FeynRules \cite{Christensen:2008py}.
\textsc{Herwig} requires more ME input and therefore has more external
interfaces, e.g. to LHA/LHEF. With the introduction of the 
\textsc{Matchbox} module this communication has become more hidden to
the user, however, since \textsc{Matchbox} contains dedicated runtime 
interfaces. Therefore normally also \textsc{Herwig} should operate 
as the top-level program. By contrast, external input is more visible in 
\textsc{Pythia}. There is no built-in ME generator or NLO subtraction
code, so whatever is not in the hardcoded library of LO processes has to 
be obtained from elsewhere, most commonly with LHA/LHEF input. The 
hierarchy is here less important; e.g.\ it is possible both to run 
\textsc{MadGraph5\_aMC@NLO} from inside \textsc{Pythia} and the other
way around, but most common is probably to run them separately, with 
intermediate LHE files stored on disk. 

\section{Other generators}
\label{sect:other}

There are many other generators, and we have no intention to list them
all. In this section we will briefly mention a few, however, notably those
related to the three main generators.

\textsc{Ariadne} \cite{Lonnblad:1992tz} is the original dipole shower,
currently not being developed further on its own, but an important 
component in some other projects.

\begin{figure}[t]
\includegraphics[width=0.58\textwidth]{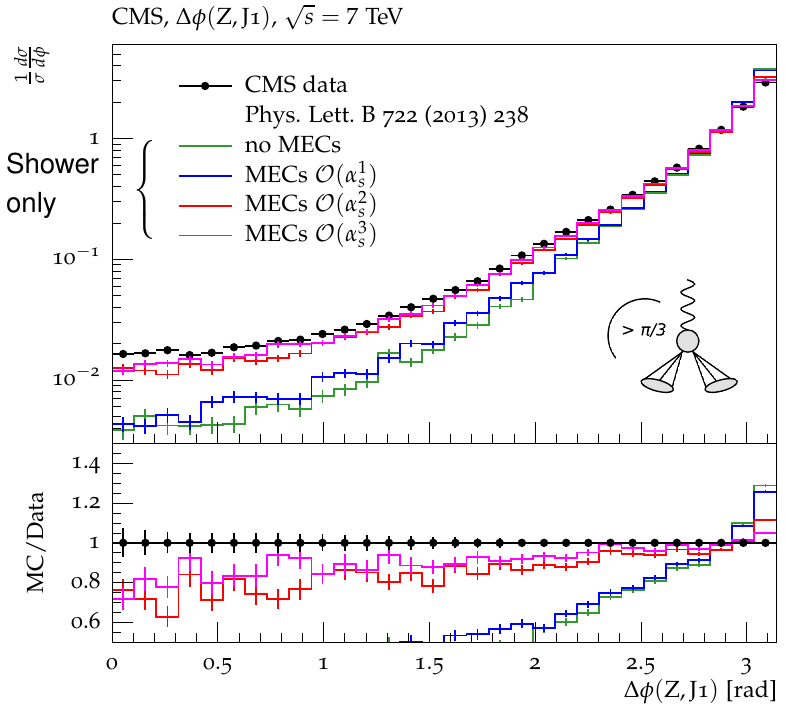}%
\hspace{0.04\textwidth}%
\includegraphics[width=0.38\textwidth]{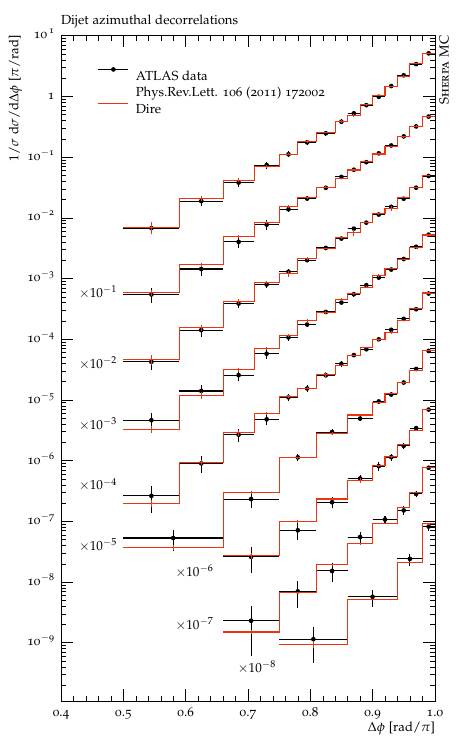}\\
\hspace*{0.3\textwidth}\textit{(a)}\hspace{0.45\textwidth}\textit{(b)}
\caption{\textit{(a)} \textsc{Vincia} results for the azimuthal angle
between a $\Zrm^0$ and the hardest recoil jet \cite{Fischer:2016vfv}. 
Agreement is further improved when hadronization is added.
\textit{(b)} \textsc{Dire} results for the dijet azimuthal decorrelation
in the \textsc{Sherpa} framework \cite{Hoche:2015sya}.}
\label{fig9}
\end{figure}

\textsc{Vincia} \cite{Giele:2007di,Giele:2011cb,Fischer:2016vfv} is a 
parton shower program, since long for FSR but now also extended to ISR. 
It is used as a plugin to \textsc{Pythia}, replacing the standard showers 
there. Branchings are defined as $2 \to 3$ processes with splitting kernels
given by antenna functions, which is a specific form of dipole expressions. 
It has a few unique features. One of them is that it is formulated 
as a Markovian process, i.e. it retains no memory of the path taken to 
reach a certain partonic configuration when deciding what to do next.
Another is that, after each trial emission, an ME correction factor is 
applied to bring it into agreement with the full (Born) ME expression to 
that order (to as high an order as feasible, and of course apart from 
Sudakov factors). Also, it allows emissions unordered in the $p_{\perp}$ 
evolution variable, although with a dampened rate. That way the full 
phase space is covered, without any ``dead zones'', which is important 
for the ME corrections to give the desired results. An example how the 
ME corrections build up order by order is shown in 
figure~\ref{fig9}\textit{a}. 

\textsc{Dire} \cite{Hoche:2015sya} is unusual in that it is one algorithm 
implemented as two completely separate codes, one for \textsc{Sherpa} and 
one for \textsc{Pythia}. This has allowed a level of technical checks 
quite unusual among event generators. (One could wish for more of this,
but of course it comes at a price that usually one cannot afford.)
It is labelled as a ``midpoint between dipole and parton showers'', in
that it uses the dipole language of $2 \to 3$ branchings, while the 
splitting functions single out radiation collinear to individual partons. 
The evolution variable is, on the other hand, based on transverse momentum 
in the soft limit, and is symmetric in emitter and spectator momentum.
The soft terms of the kernels are made less singular by the replacement
$1/(1-z) \to (1-z)/((1-z)^2 + p_{\perp}^2/M^2)$, where $M$ is the dipole mass.  
The algorithm is set up to handle negative splitting functions, e.g.\
from negative PDFs. In the future an extension to NLL is intended,
requiring new $1 \to 3$ splitting kernels and negative corrections to
the $1 \to 2$ ones (cf.\ \textsc{NLLjet} \cite{Kato:1990as}), so then 
negative contributions will become more important. An example of 
comparisons with data is shown in figure~\ref{fig9}\textit{b}. 

\textsc{Deductor} \cite{Nagy:2007ty,Nagy:2014mqa,Nagy:2015hwa}
is another parton-shower program, with a strong emphasis on the 
handling of quantum interference effects, notably in colour and spin.
It uses a dipole language, but one where all final partons share the 
recoil of a branching, whether ISR or FSR. The ordering variable is not 
a normal $p_{\perp}$ one but based on shower time arguments, which gives
a behaviour more like $|p^2 - m^2|/E$, where the numerator is the parton 
virtuality.  

\textsc{HEJ} (High Energy Jets) \cite{Andersen:2011hs} provides an 
all-orders description of processes with more than two jets. It is 
based on an approximation that should capture hard, wide-angle QCD
radiation. As such it is complementary to traditional showers, 
where the focus is more on soft and collinear emissions. Traditionally
it has been combined with the \textsc{Ariadne} shower, but efforts are
underway to interface it to \textsc{Herwig} and \textsc{Pythia}.

\textsc{Geneva} \cite{Alioli:2013hqa,Alioli:2016wqt} contains analytic
resummation of (up to) NNLL$'$ corrections to NNLO cross sections,   
using jet resolution criteria such as $N$-jettiness \cite{Stewart:2010tn}
to split the cross section into event classes. It is interfaced to 
\textsc{Pythia}.

There are also many other generators intended to describe heavy-ion physics
and/or cosmic-ray cascades in the atmosphere, such as \textsc{Hijing}
\cite{Gyulassy:1994ew}, \textsc{DPMjet} \cite{Engel:1995yda}, 
\textsc{QGSjet-II} \cite{Ostapchenko:2004ss}, and \textsc{Sybill} 
\cite{Riehn:2015oba}. These are all focused on QCD physics aspects,
i.e.\ are not ``general-purpose''. It would carry too far to describe 
them in further detail, but a few are especially interesting also for 
$\prm\prm$ applications, and are briefly mentioned here. 

\textsc{Epos} \cite{Pierog:2013ria} is based on a two-component model. 
For the collision core a collective hadronization model is used, wherein 
cluster decays are arranged to produce radial flow, and free parameters 
are adjusted to be consistent with heavy-ion data, e.g.\ for the particle
composition, thus representing a model for quark--gluon plasma
hadronization. For the collision corona instead a stringlike hadronization
model is used, akin to traditional $\prm\prm$ models. The relative 
importance of the two components is set event by event based on the
collision dynamics. \textsc{Epos} describes many aspects of LHC 
soft-physics data better than any of the general-purpose generators,
notably collective-flow and QGP-like ones (see further below). 

\textsc{Dipsy} \cite{Avsar:2006jy,Flensburg:2011kk} is a program for 
dipole evolution in transverse coordinate space, notably for the 
evolution in the initial state, leading up to the subsequent MPIs.
It can be applied both to $\prm\prm$, $\prm$A and AA collisions.
For the hadronization, an interesting extension is the formation and 
fragmentation of colour ropes \cite{Bierlich:2014xba}, hypothetical
objects created when several simple strings combine to form fields of 
a higher colour representation than the normal triplet one. One 
disadvantage is that \textsc{Dipsy} is slow to run, especially for 
$\prm$A and AA, so as an alternative a simpler wounded-nucleon model 
in the \textsc{Fritiof} \cite{Andersson:1986gw} spirit is also made 
available \cite{Bierlich:2016smv}. 

\section{Summary and outlook}

Event generators have come a long way over the last forty years,
from being a novelty met with scepticism to becoming a focal point
in much of current-day phenomenological and experimental activities.

The main theme of generator development over the last fifteen years
has been to exploit the increasing ME calculational capability, both 
in terms of more legs and of more loops. The purely perturbative 
techniques are still limited to a few partons and an even smaller
number of loops, and in particular do not fully (or at all) take into 
account the virtual corrections that are absolutely essential for a 
physical description. These are the effects that, to some approximation,
are encoded in the Sudakov factors of parton showers. To match and merge
matrix elements with parton showers thus is not only a matter of 
improving the showers in the region of hard emissions, it is also  
a matter of saving the MEs in the soft region where they do not make 
physical sense on their own. The field is teeming with ideas how to achieve
the perfect marriage between MEs and showers. Progress is steadily 
being made, but we probably are still far from a consensus.

Closely related with this trend, but also relevant on its own merits,
is the continued evolution of parton showers. One recent example is
the semiautomatic generation of scale-choice uncertainty bands.
In the future we are likely to see more emphasis on aspects that go 
beyond the simple picture of ``improved leading log'' $1 \to 2$ or 
$2 \to 3$ branchings. This has already begun, e.g.\ by studies of 
effects subleading in colour, but will involve many further aspects, 
such as $\mathcal{O}(\alpha_s^2)$ branching kernels to achieve NLL 
precision.

Somewhat forgotten in this evolution has been soft-physics aspects
of the event generation, such as multiparton interactions, colour
reconnection, diffraction and hadronization. For a long time we believed 
in jet universality, i.e.\ that nonperturbative parton fragmentation 
basically is the same in $\erm^+\erm^-$ and $\prm\prm$ collisions. This 
dream has been shattered by LHC data. Most obviously by the observation 
of the ``ridge effect'' and other manifestations of collective flow
\cite{Khachatryan:2010gv,Aad:2015gqa,Khachatryan:2016txc}. But even 
more damning is the increasing fraction of strangeness production for 
high-multiplicity $\prm\prm$ events, in a trend that nicely lines up
$\prm\prm$, $\prm$A and AA data \cite{Adam:2016emw}. Does this imply the 
(partial) formation of a quark--gluon plasma in high-multiplicity 
$\prm\prm$ collisions, contrary to the standard QGP gospel of a need for
larger volumes and time scales than $\prm\prm$ can offer? This is the 
assumption in \textsc{Epos}, which 
describes many of these features. Also the colour rope model of 
\textsc{Dipsy} looks promising. So do either of these offer the correct
answer, or what else could be going on? These are burning issues that 
are not yet reflected in the three standard generators. Clearly much 
work lies ahead of us in this area, and of a rather less transparent 
character than in the M\&M game.

Looking ahead, there is the hope that LHC will find new physics, and 
that the top priority of generator development will be to provide full
support for the study of the new processes. If not, it is likely that 
generator activities will diversify, not only to address technical 
higher-order precision but also towards a reinvigorated exploration of 
soft physics.

\section*{Acknowledgments}

Work supported in part by the Swedish Research Council, contract number 
621-2013-4287, in part by the MCnetITN FP7 Marie Curie Initial Training 
Network, contract PITN-GA-2012-315877, and in part by the the European 
Research Council (ERC) under the European Union's Horizon 2020
research and innovation programme, grant agreement No 668679.

\end{document}